\documentclass[aps,pre,twocolumn,groupedaddress,longbibliography,letterpaper]{revtex4-2}

\usepackage{amssymb}
\usepackage{amsmath}
\usepackage{graphicx}
\usepackage{hyperref}
\usepackage[caption=false]{subfig}
\usepackage{xcolor}
\usepackage[normalem]{ulem}

\newcommand\unit[1]{\ensuremath{\ \mathrm{#1}}}
\newcommand\abs[1]{\ensuremath{\lvert#1\rvert}}

\newcommand\ket[1]{\ensuremath{|#1\rangle}}

\newcommand{\derivatived}{\mathrm{d}}
\newcommand{\kB}{k_\mathrm{B}}
\newcommand{\Rb}{^{87}\mathrm{Rb}}

\newcommand{\Done}{D_1}  
\newcommand{\Py}{P_y}
\newcommand{\Pz}{P_z}
\newcommand{\Ppump}{P_{\mathrm{p}}}

\newcommand{\Praman}{P_{\mathrm{R}}}
\newcommand{\Bz}{B_z}
\newcommand{\paramvalues}{\mathbf{X}}

\newcommand{\ODpeak}{\mathrm{OD}}
\newcommand{\Ntotal}{N}
\newcommand{\NBEC}{N_\mathrm{BEC}}
\newcommand{\Nthresh}{N_1}
\newcommand{\temperature}{T}
\newcommand{\sequencetime}{t}
\newcommand{\collisionrate}{\nu_\mathrm{c}}

\newcommand{\PSD}{\mathrm{PSD}}
\newcommand{\PSDc}{\mathrm{PSD}_\mathrm{c}}
\newcommand{\nclassical}{n_\mathrm{c}}
\newcommand{\nclassicalpeak}{n_\mathrm{cp}}
\newcommand{\lambdadB}{\lambda_\mathrm{dB}}
\newcommand{\cost}{C}  
\newcommand{\costmeasured}{\cost_\mathrm{m}}  
\newcommand{\costpredicted}{\cost_\mathrm{p}}  
\newcommand{\boltzmannfactor}{f_\mathrm{B}}
\newcommand{\partitionfunction}{Z}
\newcommand{\trappotential}{U}
\newcommand{\positionvec}{\mathbf{x}}
\newcommand{\trapbottom}{\positionvec_0}
\newcommand{\crosssection}{\sigma}
\newcommand{\vrelativerms}{v_\mathrm{rms}}
\newcommand{\reducedmass}{\mu}
\newcommand{\mass}{m}
\newcommand{\omegabar}{\bar{\omega}}
\newcommand{\beamwidth}{w}

\newcommand{\beamwaisti}{\beamwidth_{i, 0}}
\newcommand{\zrayleigh}{z_\mathrm{R}}
\newcommand{\omegaradiali}{\omega_{i, \mathrm{r}}}
\newcommand{\vexpansion}{\bar{v}}

\definecolor{green}{rgb}{.5,1,.5}

\begin{document}

\title{Machine-learning-accelerated Bose-Einstein condensation}

\author{Zachary Vendeiro}
\author{Joshua Ramette}
\author{Alyssa Rudelis}
\author{Michelle Chong}
\author{Josiah Sinclair}
\author{Luke Stewart}
\author{Alban Urvoy}
\altaffiliation{Current affiliation: Laboratoire Kastler Brossel, Sorbonne Universit\'e, CNRS, ENS-Universit\'e PSL, Coll\`{e}ge de France, 4 place Jussieu, 75005 Paris, France}
\author{Vladan Vuleti\'c}
\email{vuletic@mit.edu}
\affiliation{Department of Physics, MIT-Harvard Center for Ultracold Atoms and Research Laboratory of Electronics, Massachusetts Institute of Technology, Cambridge, Massachusetts 02139, USA}

\date{\today}

\begin{abstract}
Machine learning is emerging as a technology that can enhance physics experiment execution and data analysis.
Here, we apply machine learning to accelerate the production of a Bose-Einstein condensate (BEC) of $\Rb$ atoms by Bayesian optimization of up to 55 control parameters.
This approach enables us to prepare BECs of $2.8 \times 10^3$ optically trapped $\Rb$ atoms from a room-temperature gas in $575 \unit{ms}$.
The algorithm achieves the fast BEC preparation by applying highly efficient Raman cooling to near quantum degeneracy, followed by a brief final evaporation.
We anticipate that many other physics experiments with complex nonlinear system dynamics can be significantly enhanced by a similar machine-learning approach.
\end{abstract}

\maketitle

\label{introduction}

Recently, researchers have begun applying machine learning techniques to atomic physics experiments, e.g., to enhance data processing for imaging~\cite{Picard2019,Ding2019,Seif2018,Ness2020,Guo2021}, determine the ground state and dynamics of many-body systems~\cite{Carleo2017,Saito2017}, or to identify phases and phase transitions~\cite{Wang2016,Carrasquilla2017,Torlai2019}.
One promising practical application of machine learning to atomic physics is in the optimization of control sequences with many parameters and nonlinear dynamics~\cite{Wigley2016,Tranter2018,Nakamura2019,Barker2020,Davletov2020,Wu2020}, and in particular to one of the workhorses of atomic physics, Bose-Einstein condensates (BECs)~\cite{Wigley2016,Nakamura2019,Barker2020,Davletov2020,Wu2020}.

With few exceptions~\cite{Chen2022}, experiments on BECs end with a destructive measurement, which requires repeated BEC preparation.
Approaches to increase the BEC production rate, and associated signal-to-noise ratio of the experiments, have generally relied heavily on hardware improvements~\cite{Stellmer2013b,Rudolph2015,Hung2008,Roy2016,Phelps2020}, or used atomic species with narrower optical transitions~\cite{Stellmer2013b,Roy2016,Phelps2020} than offered by the most widely utilized alkali atoms.
For alkali atoms, the tight confinement of atom-chip magnetic traps has enabled fast evaporation sequences, with a complex multi-layer atom-chip achieving BEC preparation times of $850 \unit{ms}$ for $4 \times 10^4$ atoms~\cite{Rudolph2015}.
Non-alkali atoms featuring narrow optical transitions can be used to reach lower temperatures in narrow-line MOTs~\cite{Stellmer2013b,Roy2016,Phelps2020}.
That approach, combined with a dynamically tunable optical dipole trap, has recently been used to prepare BECs of $2 \times 10^4$ erbium atoms in under $700 \unit{ms}$~\cite{Phelps2020}.

In this Letter, we demonstrate a complementary approach where, in a simple experimental setup with a broad-line MOT for a standard alkali atom, machine learning is leveraged to optimize a complex nonlinear laser and evaporative cooling process to quantum degeneracy.
Controlling a sequence with up to 55 interdependent experimental parameters, Bayesian optimization~\cite{Shahriari2015,Wigley2016,Tranter2018} finds parameter values which cool a gas from room temperature into the quantum degenerate regime in $575 \unit{ms}$, creating a BEC containing $\NBEC=2.8 \times 10^3$ atoms.
To our knowledge, this is the fastest BEC creation to date.
We identify some of the physical strategies discovered by the algorithm, and also investigate how the choice of cost function impacts the trade-off between final atom number and the purity of the created BEC.

\begin{figure}
    \includegraphics[width=\columnwidth]{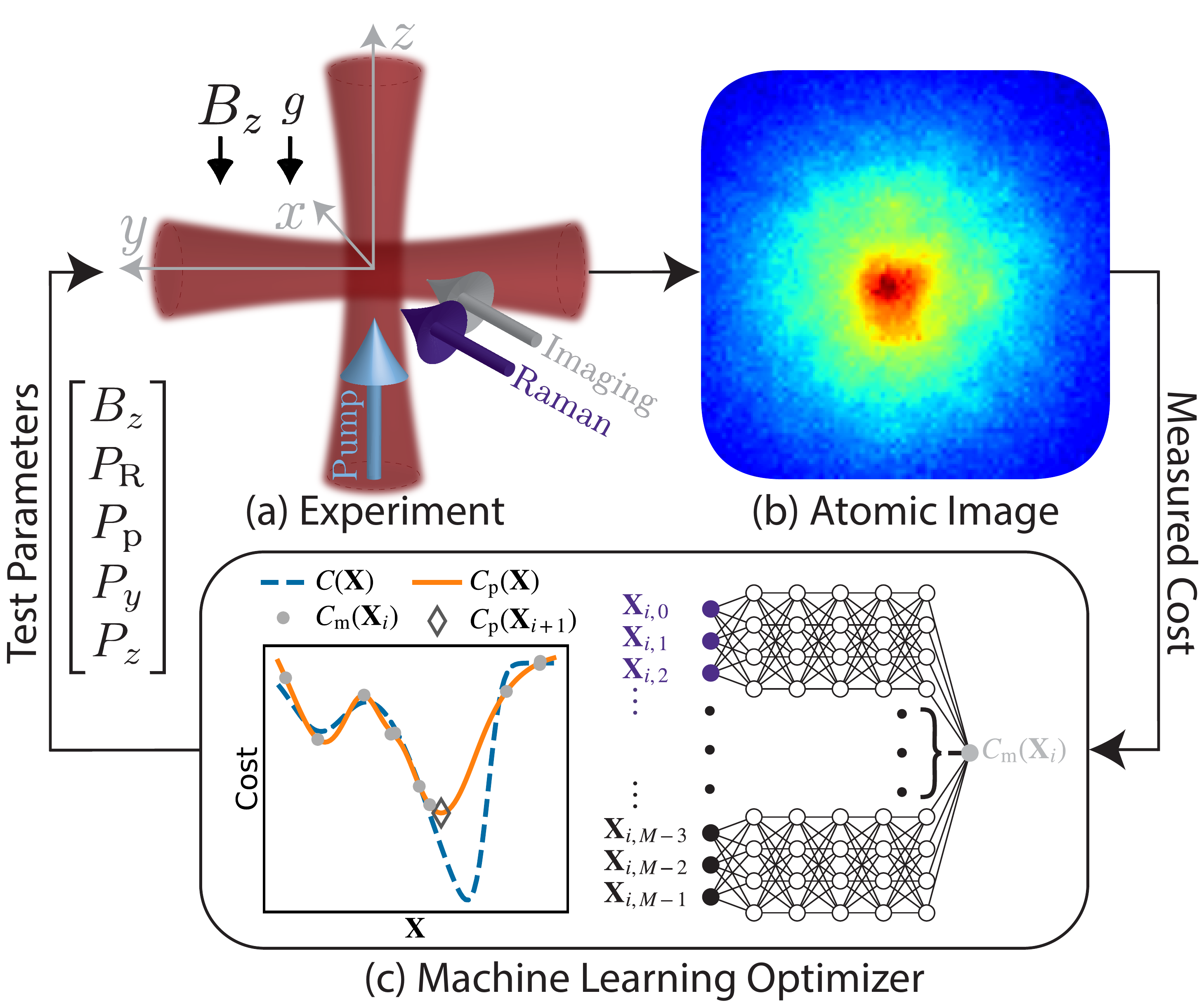}%
    \caption{
        \label{fig1}
        (a) Setup showing $1064$-nm horizontal (waist $w_h$=18$\mu$m, beam slightly tilted downward) and vertical ($w_v$=14$\mu$m) optical-trapping, $795$-nm Raman coupling ($w_R$=500$\mu$m) and optical pumping ($w_x$=30$\mu$m, $w_y \approx 1 \unit{mm}$), and $780$-nm absorption-imaging beams.
        (b) Absorption image used to extract the cost function for a set of parameter values $\paramvalues$.
        (c) Bayesian optimization with a neural network.
        The model $\costpredicted(\paramvalues)$ (orange solid line) attempts to predict the actual system performance $\cost(\paramvalues)$ (blue dashed line).
        The algorithm uses the model to predict optimal parameter values $\paramvalues_{i+1}$ (open diamond), tests those values, and performs a new iteration with an updated model.
    }
\end{figure}

Our apparatus employs only a single MOT directly loaded from a $\Rb$ background vapor, a crossed optical dipole trap, and two Raman cooling beams as depicted in Fig.~\ref{fig1}(a).
No Zeeman slower, two-dimensional MOT, atom chip~\cite{Rudolph2015}, dynamic trap shaping~\cite{Roy2016}, or strobing~\cite{Hutzler2017,Phelps2020} are necessary.
Using Raman cooling in a crossed optical dipole trap (cODT), a method that can reach very high phase-space density and even condensation~\cite{Urvoy2019}, the algorithm achieves a cooling slope of 16 orders of magnitude improvement in phase space density (PSD) per order of magnitude in atom loss ($\gamma=16$) up to the threshold to quantum degeneracy.
This is significantly better than the $\gamma = 7$ value we could obtain with extensive manual optimization under similar conditions~\cite{Urvoy2019}.

\label{atomic_physics_methods}

{\it Atomic physics methods.}\textemdash
The Raman cooling implementation used in this work is similar to that of Ref.~\cite{Urvoy2019}.
Cooling proceeds in a cODT formed by intersecting two non-interfering $1064 \unit{nm}$ beams, one horizontal and one vertical, see Fig.~\ref{fig1}(a).
Two $795$-nm beams drive the Raman cooling: the optical pumping beam and the Raman coupling beam.
Raman cooling~\cite{Kasevich1992} provides sub-Doppler cooling by driving velocity-selective Raman transitions between hyperfine states, here the $\ket{F=2,m_F=-2}$ and $\ket{2,-1}$ states of $\Rb$~\cite{Urvoy2019}.
The Raman transitions are non-dissipative so entropy is removed from the atomic gas in the form of spontaneously scattered photons as atoms are optically pumped back to the dark state $\ket{2,-2}$.
Light-assisted collisions, which typically prohibit laser cooling at high atomic densities, are suppressed by detuning the optical pumping light $4.33 \unit{GHz}$ to the red of the $\Done$ $F=2\rightarrow F'=2^\prime$ transition, where a local minimum of light-induced loss was observed~\cite{Urvoy2019}.

The cooling dynamics are controlled via five actuators:
(i) the horizontal $\Py$ and (ii) vertical $\Pz$ trap beam powers which set the trap depth and frequencies,
(iii) the Raman coupling beam power $\Praman$ which tunes the Raman rate, 
(iv) the power $\Ppump$ of the optical pumping beam which sets the optical-pumping rate (and also Raman rate), and
(v) the magnetic field $\Bz$ which adjusts the resonant velocity class for the Raman transition.
The cooling procedure is divided into stages during which the controls are linearly ramped, with the endpoints of each ramp constituting the optimization parameters.

\label{optimization_scheme}

{\it Optimization scheme.}\textemdash
The optimization problem can be formulated as the minimization of a cost function $C$, which maps a set of parameter values $\paramvalues \in \mathbb{R}^M$ to a corresponding cost value $C(\paramvalues) \in \mathbb{R}$, where $M$ is the number of optimization parameters.
The cost $C$ quantifies the results, and is generally a priori unknown, but can be extracted from measurements.
Bayesian optimization is well-suited for this type of problem as it can tolerate noise in the measured cost and typically requires testing fewer values of $\paramvalues$ than other optimization methods~\cite{Wigley2016,Tranter2018,Nakamura2019,Barker2020,Davletov2020,Wu2020}.

Bayesian optimization begins with collecting a training dataset by experimentally measuring the cost $\costmeasured(\paramvalues_i)$ for various values of sets of parameter values $\paramvalues_i$.
The $\paramvalues_i$ used to construct the training dataset are chosen by a training algorithm, which can implement another optimization algorithm or can select $\paramvalues_i$ randomly.
A model of the cost function is then fit to the training dataset which approximates the unknown true cost function $C(\paramvalues)$.
Although Bayesian optimization typically uses a Gaussian process for its model~\cite{Shahriari2015}, the present work uses neural networks~\cite{Tranter2018,Snoek2015}, which were chosen for their significantly faster fitting time for our typical number of optimization parameters.
Once the model is fit, a standard numerical optimization algorithm is applied to the modeled cost function $\costpredicted(\paramvalues)$ to determine which value $\paramvalues_{i+1}$ for the next iteration is predicted to yield the minimal cost, as depicted in Fig.~\ref{fig1}(c).
Optionally this numerical optimization can be constrained to a trust region (a smaller volume of parameter space centered around the $\paramvalues_i$ which yielded the best cost measured thus far).
The predicted optimal value $\paramvalues_{i+1}$ is then tested by experimentally measuring the corresponding cost $\costmeasured(\paramvalues_{i+1})$.
The next iteration begins by retraining the model with the new result, and making a new prediction for the optimal value of $\paramvalues$ with the updated model.
The algorithm iterates until it reaches a termination criterion, such as a set maximum number of iterations, or a set number of consecutive iterations that fail to return better results.
All optimization in this work was performed with the open-source packages M-LOOP~\cite{Wigley2016,Tranter2018} to implement the Bayesian optimization and Labscript~\cite{Starkey2013} for experimental control.
Additional implementation details are included in Appendix~\ref{bayesian_optimization_implementation}.

\label{cost_function}

{\it Cost function.}\textemdash
Since the optimization transitions the gas from the classical into the quantum degenerate regime, the final state of the gas depends strongly on how the cost function is chosen as a combination of the two experimentally accessible parameters: atom number $N$ and temperature $T$.
The classical phase space density $\PSDc$ is defined as $\PSDc \equiv \nclassicalpeak \lambdadB^3$, where $\lambdadB$ is the thermal de Broglie wavelength and $\nclassicalpeak$ is the calculated peak number density neglecting bosonic statistics (See Appendix~\ref{calculations_of_atomic_gas_properties} for calculation details).
The value of $\PSDc$ is nearly equal to the true $\PSD$ when $\PSD \ll 1$, while at the threshold to condensation, $\PSDc \sim 1$.
Since the temperature $T$ is more difficult to determine in the quantum degenerate regime, and also requires a fit to the data with potential convergence problems, we instead measure $N$ and the peak optical depth $\ODpeak$ in an absorption image.
Generally ensembles with larger $\PSDc$ have a larger atom number $\Ntotal$ and less expansion energy, which leads to a larger peak optical depth $\ODpeak$ for a given $\Ntotal$.
Guided by this, we explored cost functions of the form
\begin{equation}
    C(\paramvalues) \propto - f(\Ntotal / \Nthresh) \ODpeak^3 \Ntotal^{\alpha - 9/5},
\end{equation}
where $f(\Ntotal / \Nthresh)$ is a smoothed Heaviside step function with $\Nthresh$ chosen near the detection noise floor (see Appendix~\ref{bayesian_optimization_implementation}).
The parameter $\alpha$ in the cost function tunes the trade-off between optimizing for larger atom number or lower temperature.
For a pure BEC after sufficient time-of-flight (TOF) expansion, $\abs{C/f}$ scales as $(\NBEC)^\alpha$ (see Appendix~\ref{cost_scaling}).
For a thermal cloud, $\abs{C/f}$ is proportional to $\PSDc$ when $\alpha = -1/5$, although that value of $\alpha$ is unsuitable for condensation as increasing the atom number in the BEC requires $\alpha>0$.

\label{optimization_procedure}

{\it Optimization procedure.}\textemdash
The sequence begins with a separately optimized $99$-ms long MOT loading and compression period.
The trap beam powers are ramped to their initial Raman cooling values during the last $10 \unit{ms}$ of the MOT compression and then the magnetic field is adjusted to its initial Raman cooling value in $1 \unit{ms}$, at which point the horizontal dipole trap holds typically $\Ntotal=2.7 \times 10^5$ atoms.
We then added $100$-ms stages of Raman cooling one by one and optimized them individually.
After five stages, the algorithm tended to turn off the Raman cooling by turning down $\Ppump$ or $\Praman$ or by tuning the magnetic field $\Bz$ such that the Raman transition became off-resonant.
We then added up to six shorter $30$-ms long stages in which the optical pumping and Raman coupling beams were turned off, and the algorithm performed evaporative cooling.
Due to the reduced number of parameters, we were able to optimize the evaporation stages simultaneously, which produced a BEC.
Subsequently we shortened the Raman cooling and evaporation stages with parameter values fixed until only a small and impure BEC was produced, and then ran a global reoptimization.
In this global optimization stage, all 42 of the Raman cooling and evaporation parameters were reoptimized simultaneously using the previous values as the initial guess for $\paramvalues$.
Often a trust region set to one tenth of the allowed range for each parameter was used.
This kept the optimizer focused in regions of parameter space which produced a measurable signal, as adjusting even a single parameter too far would often result in the loss of all atoms.
We repeated this sequence shortening and reoptimization procedure until the algorithm failed to find parameters that could produce sufficiently pure BECs.

\begin{figure}
    \includegraphics[width=\columnwidth]{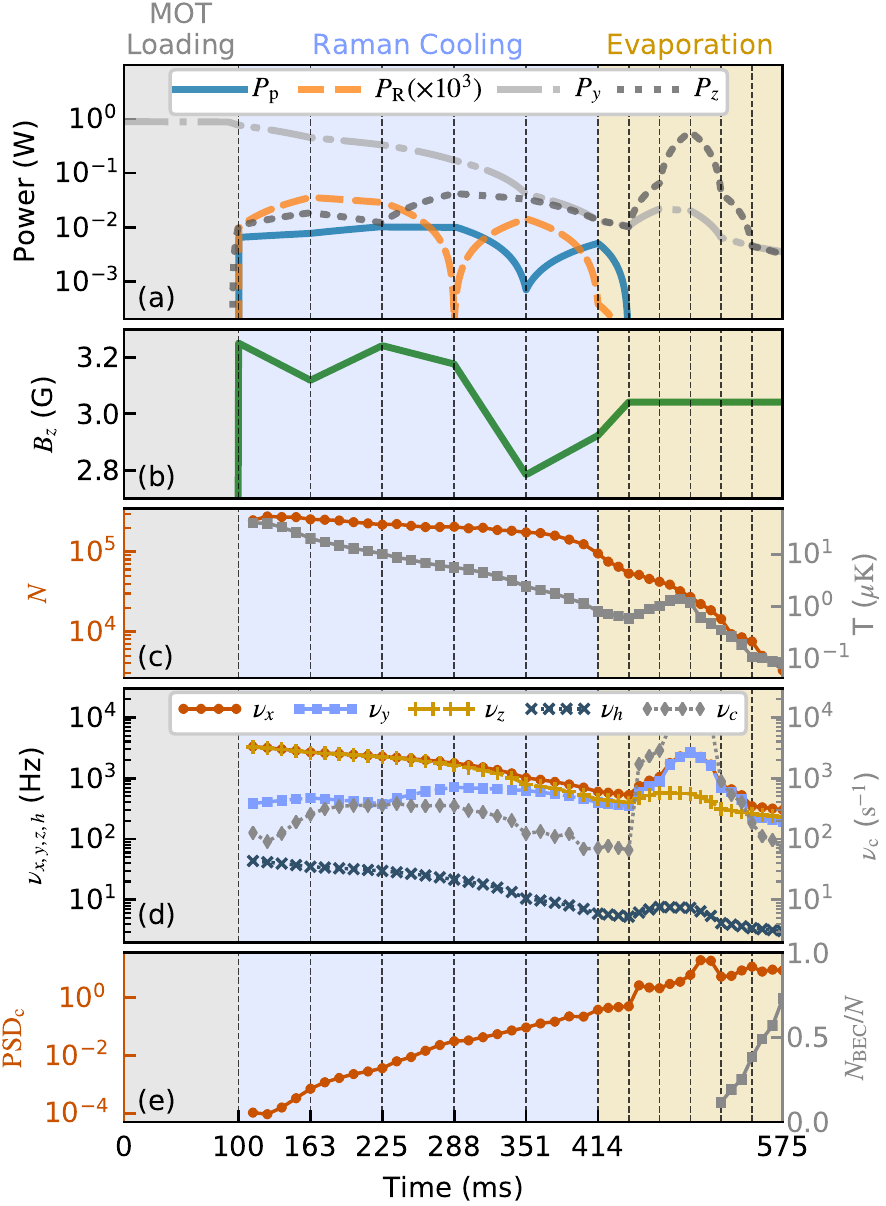}%
    \caption{
        \label{fig_sequence}
        Control waveforms (a-b) and measured trap and atomic-gas properties (c-e) of the optimized sequence.
        Gray, blue and oranges shadings mark the MOT loading, Raman cooling, and evaporation periods, respectively.
        The Raman beam power has been multiplied by $10^3$ for better visibility.
        $\nu_x, \nu_y, \nu_z, \nu_h$ are the trap vibrations frequencies in the $x, y, z$ directions and in the horizontal trap, respectively; $\collisionrate$ is the atomic collision rate.
        $\PSDc$ does not account for bosonic statistics and changes slowly while the BEC forms quickly above threshold.
        Calculations assume thermal equilibrium.
        }
\end{figure}

The required beam powers generally varied over several orders of magnitude, so the logarithm of their powers were used as entries in $\paramvalues$, while the magnetic-field control parameter $\Bz$ was kept a linear parameter.
A feedforward adjustment was included in the $\Bz$ control values to account for the light shift of the $\ket{2,-1}$ state by the optical pumping beam.
We averaged over five repetitions of the experiment for each set of parameter values tested.
The number of iterations per optimization varied but was typically ${\sim} 1000$ (including the initial training), and required several hours, both for the single-stage optimizations and the full-sequence optimizations.
A simpler optimization procedure was also attempted which did not involve optimizations of individual stages.
Instead the sequence was divided into ten $100 \unit{ms}$ stages and all 55 parameters were optimized from scratch simultaneously.
That approach combined with the shortening and reoptimizing procedure successfully produced a similar BEC, albeit in slightly longer time ($650 \unit{ms}$ vs $575 \unit{ms}$), possibly due to the optimization becoming trapped in a local optimum (see Appendix~\ref{bayesian_optimization_implementation} for further discussion).

\label{results}

\begin{figure}
    \includegraphics[width=\columnwidth]{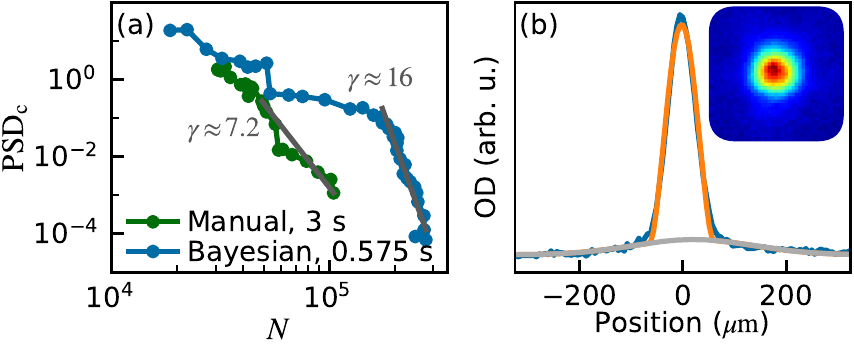}%
    \caption{
        \label{fig_performance}
        Results of the $575 \unit{ms}$ optimized sequence.
        (a) $\PSDc$ vs atom number $\Ntotal$.
        Initial cooling until \mbox{$\PSDc \sim 10^{-1}$} is very efficient with $\gamma \approx 16$ (gray line).
        The performance of the much slower (3~s-long) manually optimized sequence of Ref.~\cite{Urvoy2019} is shown for comparison ($\gamma \approx 7$).
        (b) Cross section of 24-ms TOF image (inset) shows a BEC (orange fit) with small thermal wings.
    }
\end{figure}

\label{results_and_physical_interpretation}

{\it Results and physical interpretation.}\textemdash
The best discovered $575$-ms long control sequence and corresponding results are depicted in Fig.~\ref{fig_sequence} and Fig.~\ref{fig_performance}.
Notably, the algorithm discovered gray molasses~\cite{Weidemuller1994,Boiron1995} in the MOT phase, which it applies at the end of the compression sequence.
This outperforms the bright molasses~\cite{Lett1988,Dalibard1989} that was previously used in the manually optimized compression sequence, with the gray molasses loading a similar number of atoms ten times faster.
After the MOT loading stage and transfer into the cODT, five ${\sim} 63$-ms long stages of Raman cooling follow, and then the optical pumping and Raman beams are ramped off, followed by six ${\sim} 27$-ms long evaporation stages.
As observed in previous work~\cite{Tranter2018,Nakamura2019,Barker2020}, the ramps produced by Bayesian optimization are non-monotonic and appear non-intuitive, but they outperform the routines we found by manual optimization.
A reason for the non-monotonic waveforms may be that the cost function includes many local minima.
The optimization can settle into any one of these local optima randomly, and produce complex but specific waveforms, as observed in Ref.~\cite{Tranter2018}.
Despite the non-monotonic ramps, $\PSDc$ increases smoothly exponentially during this part of the sequence (Fig.~\ref{fig_sequence}(e)), due in part to the finite thermalization rate.

By shortening the sequence we are asking the algorithm to maximize the cooling speed, which is limited by the lower of the collisional rate $\collisionrate$ and the trap vibration frequencies $\nu_{x,y,z}$ \cite{Vuletic1999a}.
When the gas is still hot, we have $\collisionrate \ll \nu_{x,y,z}$, and the algorithm employs Raman cooling to increase the density and collision rate (Fig.~\ref{fig_sequence}(d)).
However, when $\collisionrate$ approaches the lowest trap vibration frequency $\nu_y$ near the time $\sequencetime = 225$~ms, the algorithm starts to reduce the Raman rate, and a little later the optical pumping rate, in order to reduce light induced collisions that scale with $\collisionrate$, rather than the trap vibration frequency.
Subsequently, for times $\sequencetime > 225$~ms, the cooling proceeds near optimally, with the collision rate close to, but a little below, the trap vibration frequencies.
Furthermore, as the system approaches condensation near $\sequencetime = 410$~ms, the collision rate is somewhat lowered to reduce light-induced atom loss (Fig.~\ref{fig_sequence}(c)).

Another effect limiting the cooling speed is the loading of the atoms from the single horizontal trap, in which the sample is initially prepared, into the crossed dipole trap (see the movie in SM~\cite{SuppMat}).
Initially, the vertical-beam power $\Pz$ is held low to avoid creating a high-density dimple region which would lead to excess loss during Raman cooling.
Later, $\Pz$ is ramped up to gather atoms from the horizontal trap beam into the overlap region of the cODT in order to increase the collision rate and speed up evaporative cooling.
The relatively sudden ramping of the trap power up and then back down visible in Fig.~\ref{fig_sequence}a likely involves an optimal-control-like process since the trap compression and relaxation are faster than the axial period of the horizontal trap of $\sim 200$~ms.

The optimization tended to turn off the Raman cooling after five stages because the cloud temperature $\temperature$ was below the effective recoil temperature~\cite{Urvoy2019} where Raman cooling, even with optimal parameters, becomes too slow, while leading to trap loss and heating due to light-assisted collisions~\cite{Burnett1996}.
The Bayesian optimization recognized this and shut down the Raman cooling at this point, with the atomic gas close to condensation.
Subsequently, at higher compression which is primarily achieved by increasing the vertical beam power, the horizontal trap power is reduced and atoms are efficiently evaporated along the direction of gravity in the tilted potential~\cite{Hung2008} (see the movie in the SM~\cite{SuppMat}).
Note also that once the atoms have been loaded into the crossed-trap region (after $\sequencetime = 350$~ms), the algorithm makes all trap vibration frequencies similar, which provides the fastest overall thermalization, and hence largest cooling speed.

The BEC is fully prepared at the end of the evaporation stages, $575 \unit{ms}$ after the start of the MOT loading.
The final cloud contains $3.7 \times 10^3$ total atoms and is shown in Fig.~\ref{fig_performance}(b).
A bimodal fit of the cloud indicates that $2.8 \times 10^3$ atoms (76 \%) are in the BEC.
Although the sequence was optimized for speed rather than efficiency, the initial cooling occurs with a logarithmic slope $\gamma = \derivatived (\log \PSDc) / \derivatived(\log \Ntotal) \approx 16$.

\label{cost_function_impact}

\begin{figure}
    \includegraphics[width=\columnwidth]{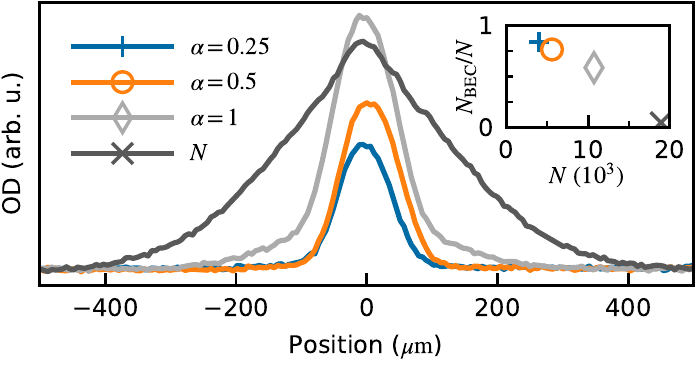}%
    \caption{
        \label{cost_function_comparison}
        Cross sections of 24-ms TOF images (200 averages) optimized for different cost function parameter $\alpha$ (see main text) with 1-s long sequences, demonstrating the trade off between optimizing for atom number or temperature.
        Also plotted are the results of optimizing for atom number $\Ntotal$ only.
        Inset: Condensate fraction $\NBEC/\Ntotal$ vs $\Ntotal$ for different $\alpha$.
    }
\end{figure}

{\it Cost function impact.}\textemdash
The atomic gases produced by sequences optimized for different values of $\alpha$ are presented in Fig.~\ref{cost_function_comparison}, as well as the results when optimizing for total atom number ($\Ntotal$).
Larger values of $\alpha$ result in more atoms, but at higher temperature and lower condensate fraction, while smaller values of $\alpha$ produce purer BECs, but with fewer atoms overall.
Setting $\alpha$ to $0.5$ was found to make a reasonable compromise (orange curve in Fig.~\ref{cost_function_comparison}); so that value was used for the final full-sequence optimization which yielded the data presented in Fig.~\ref{fig_sequence}.

\label{outlook}

{\it Outlook.}\textemdash
In conclusion, we have demonstrated that Raman cooling with far detuned optical-pumping light combined with a final evaporation can rapidly produce BECs with a comparatively simple apparatus, even with a standard alkali atom which lacks narrow optical transitions.
Bayesian optimization greatly eased the search for a short sequence to BEC, quickly discovering initially unintuitive yet high-performing sequences.
Inspection of the parameters chosen by the algorithm reveals several physical strategies, such as adjusting a collision rate close to, but below the trap vibration frequencies to maximize the thermalization and cooling speed while minimizing density-dependent atom loss, non-adiabatic loading into the crossed-trap dimple, and creating a nearly isotropic trap for efficient evaporation.
In future applications, faster condensation can likely be achieved by including dynamical tuning of trap size~\cite{Roy2016}, while user intervention may be further reduced by factoring the sequence length into the assigned cost~\cite{Barker2020}.
We anticipate that many other experimental procedures in atomic physics and beyond can be improved by machine learning.

\begin{acknowledgments}
The authors would like to thank Martin Zwierlein for inspiring physics discussions, and Michael Hush, Harry Slatyer, Philip Starkey, Christopher Billington, and Russell Anderson for stimulating discussions and software assistance.
This work was in part supported by the NSF, NSF CUA, NASA, DoE, and MURI through AFOSR.
\end{acknowledgments}

\appendix

\section{Bayesian Optimization Implementation}
\label{bayesian_optimization_implementation}

In M-LOOP's implementation of Bayesian optimization, the training algorithm used to pick parameters and generate a training dataset is also run periodically even after the training dataset is complete~\cite{Wigley2016,Tranter2018}.
In particular, once sufficient training data is acquired, three independent neural networks are trained.
Each neural net is fully connected and consists of an input layer with one node for each optimization parameter, followed by five hidden layers with 64 nodes each, and then an output layer with a single node.
Once the training has completed, each neural network is used to generate a set of parameter values $\paramvalues$ which it predicts to be optimal, and each of those three $\paramvalues$ are experimentally tested.
Then another iteration of the training algorithm is performed and the $\paramvalues$ it suggests are also tested.
The results from all four of these measurements are included in the next training of the neural nets for the subsequent Bayesian optimization iteration.
The additional iterations of the training algorithm are intended to encourage parameter space exploration and provide unbiased data~\cite{Wigley2016,Tranter2018}.

In this work, the absorption images used to measure the cost function were generally taken after 1.5 to 8 ms of time-of-flight (TOF) expansion.
We averaged over five repetitions of the experiment for each set of parameter values tested, which took ${\sim} 10 \unit{s}$ accounting for experimental and analysis overhead.
Simply taking the largest optical depth measured in any single pixel of an absorption image as $\ODpeak$ makes it prone to noise, so $\ODpeak$ was set to the average OD of several pixels with the largest OD to reduce noise.
To compare different sequences on an equal footing during optimizations, the trap beams were always ramped to a fixed power setting before releasing the atoms for TOF imaging.
This final fixed ramp is only necessary during optimizations and is omitted from the sequence once the optimizations are complete.
The smoothed Heaviside step function $f(\Ntotal / \Nthresh)$ included in the cost function ensures that the cost does not diverge at low $\Ntotal$ while having little effect when $\Ntotal$ is above the measurement noise floor.
The form of $f(\Ntotal / \Nthresh)$ is inspired by the expression for the excited state population of a two-level system in thermal equilibrium and it is defined as
\begin{equation}
    f(\Ntotal / \Nthresh) = 
    \begin{cases} 
        \left(\frac{2}{e^{\Nthresh / \Ntotal} + 1}\right) & \Ntotal > 0 \\
        0 & \Ntotal \leq 0
    \end{cases}
\end{equation}

For many of the optimizations in this work, particularly those with tens of parameters, the cost function landscape is ``sparse'' in the sense that most sets of parameter values yield poor results with a signal below the measurement noise floor.
Thus the actual performance for such $\paramvalues$ cannot be measured, and testing them provides little information to the model.
This leads to large regions of parameter space where there is no measurable signal and the direction towards better values cannot be inferred.
There are two notable consequences of this.
Firstly, for such optimizations it is generally necessary to provide initial values to the optimization which give a nonzero signal.
Without a good starting point, the training dataset will often only include measurements dominated by noise, making it exceedingly unlikely for the Bayesian optimization to succeed.
Secondly, for such optimizations it is generally helpful to specify a trust region.
This limits the extent of excursions as the optimizer explores parameter space, making it more likely to test parameter values which yield a measurable signal.
However, this does come at the cost that it makes it less likely for the optimizer to jump from one local minimum to another better minimum.
We often performed the same optimization with and without a trust region in parallel.
This could be done without significantly extending the duration of optimizations because the analysis for each iteration typically took longer than the time required to perform the experiment.
Thus one optimization could run experiments while the other analyzed its most recent results.
For optimizations with many parameters, the results with a trust region were typically as good as or better than those without.
This is likely a consequence of that fact that, given the sparsity of the cost function landscape, it is unlikely for the optimizer to discover another local optimum.
Thus it is better for the optimizer to focus on modeling the region of parameter space around the local optimum rather than fruitlessly searching for another local optimum.

\begin{figure*}
    \includegraphics[width=\textwidth]{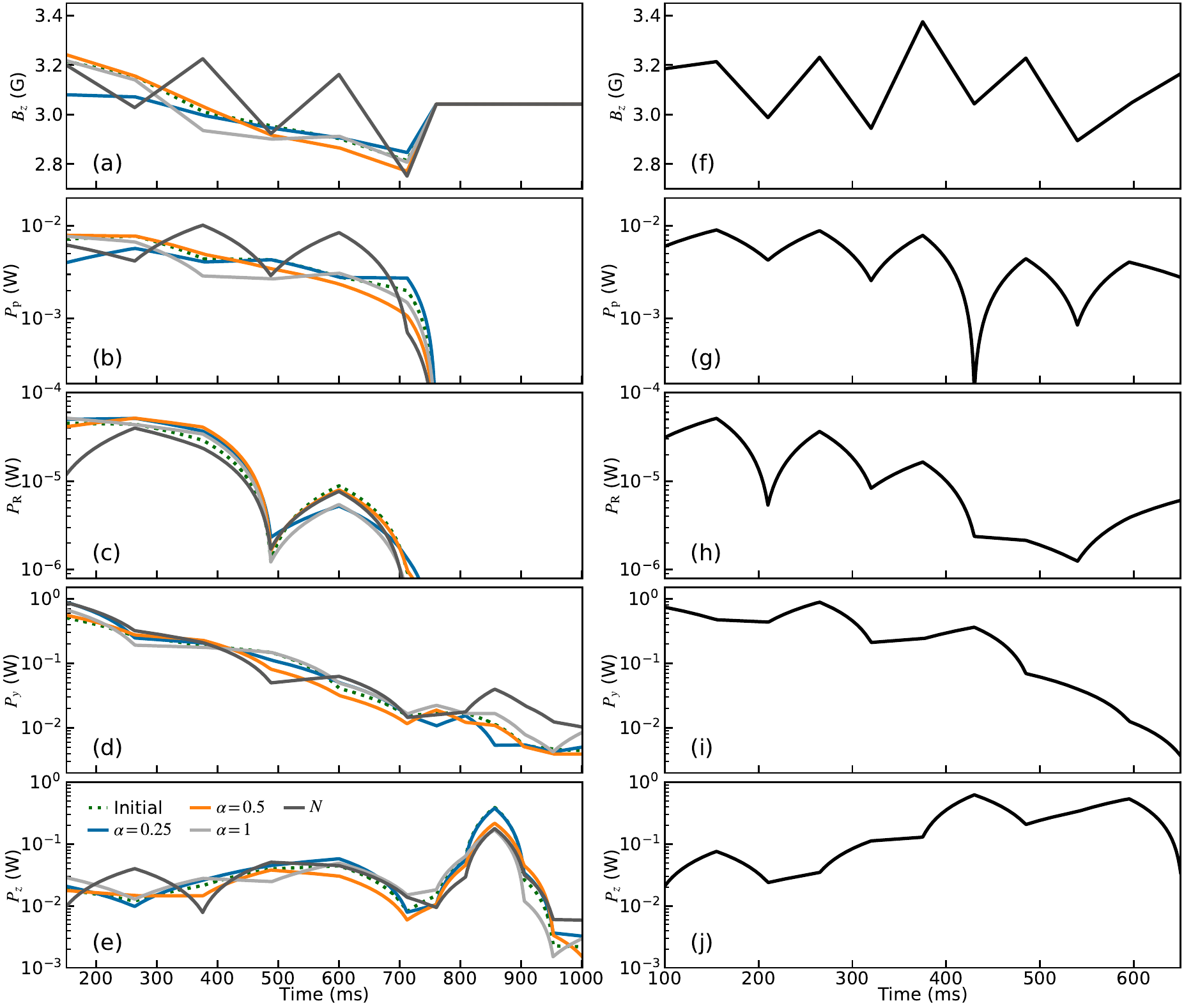}%
    \caption{
        \label{many_optimized_waveforms}
        (a) The control waveforms for the 1 second sequences corresponding to the results presented in Fig.~\ref{cost_function_comparison}, as well as the initial waveform used as the starting point for each of those optimizations.
        (b) The control waveforms for the $650 \unit{ms}$ 10-stage sequence optimized from scratch rather than stage-by-stage initially, which was optimized with $\alpha = 0.5$.
        Note the differing limits for the $x$-axes between (a) and (b).
        The sequences in (a) include a $50 \unit{ms}$ magnetic coil ramp duration which was later reduced to $1 \unit{ms}$.
        The MOT sections of the sequences have been omitted for simplicity.
        Note that the waveforms in (a) are mostly qualitatively similar despite being optimized for different cost functions.
        The waveforms in (b) are more qualitatively distinct from those in (a), even for the orange curves which were also optimized with $\alpha = 0.5$.
        This suggests that the independent optimizations likely become trapped around disparate local minima.
        On the other hand, tuning the cost function while providing the same initial parameter values each time typically causes only smaller deviations around the initial values.
    }
\end{figure*}

The sparsity of the cost landscape and necessity for providing initial parameters which produce measurable results posed a difficulty when we optimized an entire sequence from scratch at once (rather than initially adding one cooling stage at a time).
We resolved this by reducing the time of flight to $1.5 \unit{ms}$ for the first optimization.
With such a short time of flight, even poor parameter values could produce clouds with a peak optical depth above the measurement noise floor.
Due to the finite dynamic range of the absorption imaging, the results of this first optimization produced a cloud which saturated the measurement and thus made it impossible to accurately quantify performance for the best-performing values.
The next optimizations were performed with the same sequence duration, but the time of flight increased to $5 \unit{ms}$ then $8 \unit{ms}$.
This made it possible to better discern differences between high-performing sets of parameter values at the cost of increasing the performance required to produce a signal above the noise floor and thus increasing the sparsity of the cost function landscape.
The procedure of shortening then re-optimizing the sequence was then applied, resulting in the sequence presented in Fig.~\ref{many_optimized_waveforms}(b), which produced a BEC in $650 \unit{ms}$.
The parameter $\alpha$ was set to $0.5$ throughout this procedure.

Although it is not strictly fair to do so due to the differing parameterizations, it is still informative to compare the control waveforms of the independently-optimized $650 \unit{ms}$ sequence to those from Fig.~\ref{cost_function_comparison}.
These waveforms are presented in Fig.~\ref{many_optimized_waveforms}.
The sequences of Fig.~\ref{cost_function_comparison} optimized for different $\alpha$ all had fairly similar waveforms.
On the other hand, the $650 \unit{ms}$ sequence had a qualitatively different waveform.
For example, it lacks the sudden rise and drop in vertical trap power towards the end of the sequence present in the other waveforms.
This suggests that it has converged to a qualitatively-different local optimum.
On the other hand, the sequences of Fig.~\ref{cost_function_comparison} were all optimized with a trust region and the same initial $\paramvalues$.
Thus those optimizations primarily performed a local search, only slightly tuning $\paramvalues$ to tailor the sequence for their particular value of $\alpha$.
Although there are small differences in parameterization, the fact that two different optimizations with the same value of $\alpha$ produce sequences that differ more than optimizations with the same initial $\paramvalues$ but different $\alpha$ supports the notion that the cost function landscape includes multiple local minima, as suggested in the main text.

\section{Calculations of Atomic Gas Properties}
\label{calculations_of_atomic_gas_properties}

The classical phase space density is defined as $\PSDc = \nclassicalpeak \lambdadB^3$ where $\nclassicalpeak$ is the peak number density calculated for a classical gas (i.e. neglecting Bosonic statistics) and $\lambdadB = h / \sqrt{2 \pi m \kB \temperature}$ is the de Broglie wavelength.
Here $h$ is the Planck constant, $m$ is the mass of an atom, and $\kB$ is the Boltzmann constant.
To calculate $\PSDc$ for a cloud, its atom number $\Ntotal$ and temperature $\temperature$ are measured and it is assumed to be in thermal equilibrium.
The value of $\lambdadB$ is easily calculated from the measured temperature.
The partition function $\partitionfunction = \int \boltzmannfactor(\positionvec) \derivatived V$ is then calculated by numerically integrating the Boltzmann factor $\boltzmannfactor(\positionvec) = \exp \left[ -\trappotential(\positionvec) / (\kB T)\right]$ over the trap volume, where $\trappotential(\positionvec)$ is the trap potential at position $\positionvec$.
The $\trappotential(\positionvec)$ is taken to be the sum of two Gaussian beams, one for each cODT beam, and gravity is neglected for simplicity.
Each Gaussian beam with peak depth $\trappotential_{i, 0}$ and waist $\beamwaisti$ contributes a potential of the form
\begin{equation}
    \trappotential_i(\positionvec) = \trappotential_{i, 0} \left( \frac{\beamwaisti}{\beamwidth_i(z^\prime)} \right)^2 \exp \left( \frac{-2 (r^\prime)^2}{ \beamwidth_i(z^\prime)^2} \right)
\end{equation}
where $\beamwidth_i(z^\prime) = \beamwaisti \sqrt{1 + (z^\prime/\zrayleigh)^2}$ is the spatially-varying beam width and $\zrayleigh = \pi \beamwaisti^2 / \lambda$ is the Rayleigh range.
The primed coordinates $z^\prime$ and $r^\prime$ are taken to be along and perpendicular to the beam's propagation direction respectively.
The value of $\nclassicalpeak$ can be calculated as $\Ntotal \boltzmannfactor(\trapbottom) / Z$ where $\trapbottom$ is the position of the bottom of the trap.
Finally $\PSDc$ is evaluated from its definition in terms of $\nclassicalpeak$ and $\lambdadB$.
Notably, for much of the sequence the atomic cloud extends out of the cODT region and into the wings of the horizontal ODT, in which case the trap potential seen by the cloud is not harmonic.
Thus the well-known result $\PSDc = \Ntotal (\hbar \omegabar)^3 / (\kB \temperature)^3$ for a harmonic trap with geometric mean trap frequency $\omegabar$ cannot be used for most of the sequence.

Calculation of the mean collision rate $\collisionrate$ requires averaging the collision rate $\nclassical \crosssection \vrelativerms$ over the cloud where $\crosssection$ is the atomic collision cross section and $\vrelativerms$ is the root-mean-square relative velocity of atoms in the cloud.
The value of $\nclassical$ varies over the trap and obeys $\nclassical(x) = \Ntotal \boltzmannfactor(x) / \partitionfunction$, again neglecting Bosonic statistics.
From equipartition for a 3D gas, $(1/2) \reducedmass \vrelativerms^2 = (3/2) \kB \temperature$ where $\reducedmass = \mass / 2$ is the reduced mass for two atoms.
Thus the value of $\vrelativerms$ is given by $\sqrt{6 \kB \temperature / \mass}$.
The local collision rate is averaged by integrating $\nclassical \crosssection \vrelativerms$ over the cloud, weighted by the 1-atom number density $\nclassical / \Ntotal$, yielding
\begin{equation}
    \collisionrate = \Ntotal \crosssection \sqrt{\frac{6 \kB \temperature}{\mass}} \int \left(\frac{\boltzmannfactor(x)}{\partitionfunction}\right)^2 \derivatived V
\end{equation}

The above calculations assume that the cloud is in thermal equilibrium, which is often a good approximation.
However, after about $440 \unit{ms}$ of the final optimized $575 \unit{ms}$ sequence, the power in the vertical trapping beam $\Pz$ is rapidly increased, as can be seen in Fig.~\ref{fig_sequence}(a).
This change is likely non-adiabatic for atoms in the wings of the horizontal ODT and the cloud may no longer be in thermal equilibrium.
This is likely why the calculated $\PSDc$ appears to increase beyond ${\sim} 1$ before the appearance of a BEC.
Notably this non-adiabatic portion of the sequence occurs only after $\PSDc$ has reached $0.4$, and thus it does not affect the cooling efficiency estimate of $\gamma \approx 16$ for the cooling up to $\PSDc = 0.1$.

The peak trap depth $\trappotential_{i, 0}$ for each beam was determined from the beam waist $\beamwaisti$ and radial trap frequency $\omegaradiali$ measured for each beam.
The beam waists, defined as the radius at which the intensity falls to $1/e^2$ of its peak value, were measured by profiling the trap beams on a separate test setup which focused the light outside of the vacuum chamber.
The trap frequencies were directly measured by carefully perturbing the position of a cloud in the cODT and observing its oscillations.
Before perturbing the cloud, it was first cooled sufficiently to make it well-confined to the central region of the cODT so that the potential was approximately harmonic.
The peak trap depth for each beam could then be calculated as $\trappotential_{i, 0} = m \omegaradiali^2 \beamwaisti^2 / 4$.
This expression can be derived by equating the spring constant for the trap in the radial direction $k = \derivatived^2 \trappotential_i(\positionvec) / (\derivatived r^\prime)^2 |_{\positionvec = \trapbottom}$ to its value for a harmonic oscillator $k = m \omegaradiali^2$.

\section{Cost Scaling}
\label{cost_scaling}

The peak optical depth $\ODpeak$ of a pure BEC after sufficient time of flight expansion scales as $\ODpeak \propto \NBEC / A$ where $A$ is the area of the cloud in the image.
The area scales in proportion to $\vexpansion^2$ where $\vexpansion$ is the expansion velocity, which is related to the BEC chemical potential via $(1/2)m \vexpansion^2 = (2/7) \mu$ in a harmonic trap~\cite{Dalfovo1999}.
Thus $A \propto \mu$.
Furthermore, the chemical potential for a harmonically-trapped BEC scales as $\mu \propto \NBEC^{2/5}$~\cite{Dalfovo1999}, so $A \propto \NBEC^{2/5}$ and $\ODpeak \propto \NBEC^{3/5}$.
The expression $\ODpeak^3 \Ntotal^{\alpha - 9/5}$ then scales as $(\NBEC)^\alpha$.
Notably this scaling also applies to a harmonically-trapped BEC when imaged in situ.
There, the BEC radius $R$ scales as $R \propto \NBEC^{1/5}$~\cite{Dalfovo1999}.
In that case, $A \propto R^2 \propto \NBEC^{2/5}$ as before.
The same arguments then apply again, indicating that $\ODpeak^3 \Ntotal^{\alpha - 9/5}$ scales as $(\NBEC)^\alpha$ for a harmonically-trapped BEC in situ just as it does for a BEC after long time of flight expansion.

The scaling of $\ODpeak^3 \Ntotal^{\alpha - 9/5}$ for a purely thermal cloud is also of note.
For a harmonically-trapped thermal cloud, the RMS size in a given direction for any time of flight is proportional to $T^{1/2}$, so $A \propto T$.
Thus $\ODpeak \propto \Ntotal / T$ and $\ODpeak^3 \Ntotal^{\alpha - 9/5}$ scales in proportion to $\Ntotal^{\alpha + (6/5)}/T^3$.
Clouds with smaller temperatures are favored by the cost function, and clouds with larger atom numbers are favored as long as $\alpha > -6/5$.
For the case $\alpha = -1/5$, the value of $\ODpeak^3 \Ntotal^{\alpha - 9/5}$ scales in proportion to $\Ntotal / T^3$, which is proportional to $\PSDc$.
That choice of $\alpha$ was often used when optimizing individual stages before reaching the threshold to BEC.
However note that this choice of $\alpha$ leads to the scaling $\ODpeak^3 \Ntotal^{\alpha - 9/5} \propto \NBEC^{-1/5}$ for a pure BEC and is thus not a good choice when the cloud reaches condensation.

\section{Raman Cooling Laser}
\label{raman_cooling_laser}

Standard Doppler cooling requires a laser with linewidth narrow compared to the optical transition linewidth in order to achieve optimal temperatures.
This places stringent technical requirements for Doppler cooling on narrow optical transitions.
By contrast, Raman cooling can achieve similar velocity resolution and associated temperatures with a comparatively broad laser.
In this work, the light for the Raman coupling and optical pumping beams, which drive the up-leg and down-leg of the Raman transition respectively, was derived from the same laser.
This ensures that any laser frequency noise is common mode between the two legs of the Raman transition and makes it possible to resolve Doppler shifts much smaller than the laser linewidth.
A DBR laser diode (Photodigm PH795DBR180TS) without an external cavity was sufficient to generate the Raman cooling light.
The forgiving laser linewidth requirements further simplify implementation of our BEC production approach compared to schemes which require Doppler cooling on narrow optical transitions.
Thus our approach may be useful even for species which include narrow optical transitions.

\bibliography{references}

\end{document}